\documentclass[a4paper,usenatbib]{mn2e} 
\usepackage{amssymb, textcomp, verbatim, graphicx, times, overpic}

\def\deg{^o}

\title{The Twisted Radio Structure of PSO\,J334.2028+01.4075, Still a
  Supermassive Binary Black Hole Candidate}

\author[Mooley et al.]{K.~P. Mooley$^{1,9}$, J.~M. Wrobel$^{2,3}$, M.~M. Anderson $^{4}$, G. Hallinan $^{4}$ \\
  $^{1}$ Centre for Astrophysical Surveys, University of Oxford, Denys Wilkinson Building, Keble Road, Oxford OX1 3RH; Email: kunal.mooley@physics.ox.ac.uk\\
  $^{2}$ National Radio Astronomy Observatory, P.O. Box O, Socorro, NM 87801, USA\\
  $^{3}$ National Science Foundation, 4201 Wilson Boulevard, Arlington, VA 22230, USA\\
  $^{4}$ California Institute of Technology, 1200 East California Boulevard, MC 249-17, Pasadena, CA 91125\\
  $^{9}$ Hintze Research Fellow}

\begin{document}
\maketitle

\begin{abstract}
  Supermassive binary black holes (BBHs) on sub-parsec scales are
  prime targets for gravitational wave experiments.  They also provide
  insights on close binary evolution and hierarchical structure
  formation.  Sub-parsec BBHs cannot be spatially resolved but
  indirect methods can identify candidates.  In 2015 Liu et
  al.\ reported an optical-continuum periodicity in the quasar
  PSO\,J334.2028+01.4075, with the estimated mass and rest-frame
  period suggesting an orbital separation of about 0.006 pc (0.7
  $\mu$arcsec).  
  The persistence of the quasar's optical periodicity has recently been disfavored over an extended baseline. 
  However, if a radio jet is launched from a sub-parsec
  BBH, the binary's properties can influence the radio structure on
  larger scales.  Here, we use the Very Long Baseline Array (VLBA) and
  Karl G.\ Jansky Very Large Array (VLA) to study the parsec- and
  kiloparsec-scale emission energized by the quasar's putative BBH.
  We find two VLBA components separated by 3.6 mas (30 pc),
  tentatively identifying one as the VLBA ``core'' from which the
  other was ejected.  The VLBA components contribute to a point-like,
  time-variable VLA source that is straddled by lobes spanning 8
  arcsec (66 kpc).  We classify PSO\,J334.2028+01.4075 as a
  lobe-dominated quasar, albeit with an atypically large twist of
  $39\deg$ between its elongation position angles on parsec- and
  kiloparsec-scales.  By analogy with 3C\,207, a well-studied
  lobe-dominated quasar with a similarly-rare twist, we speculate that
  PSO\,J334.2028+01.4075 could be ejecting jet components over an
  inner cone that traces a precessing jet in a BBH system.
\end{abstract}


\section{Introduction}\label{sec:intro}

It is widely believed that some galaxy mergers could lead to the
formation, and eventually the coalescence, of supermassive binary
black holes \citep[BBHs; reviewed by][]{colpi2014}.  If gas is
available in a galaxy merger, accretion onto its BHs may occur.  This
could produce dual active galactic nuclei (AGN) on kiloparsec-scales
that can be spatially separated and identified by, for example, their
radio, optical or X-ray emission \citep[e.g.,][]
{deane2014,comerford2015,fu2015a}.  The demographics of dual AGN are
critical for understanding the BH growth and AGN activation induced by
galaxy mergers \citep[e.g.,][]{fu2015b}.

Whether or not both BHs retain their AGN trappings, it is expected
that the separation between the BHs will, over time, shrink to the
sub-parsec scales where the BHs orbit each other in a bound system
\citep{volonteri2009}.  Further shrinkage of the orbit should
culminate in the coalescence of the BHs, with the concomittant
emission of gravitational waves \citep[reviewed by][]{schnittman2013}.
The aggregate of many coalescence events is predicted to dominate the
stochastic gravitational wave background.  Observations of that
background are underway at $\mu$Hz frequencies with pulsar timing
arrays \citep[reviewed by][]{burke-spolaor2015} or are planned at mHz
frequencies for space missions \citep[reveiwed by][]{barausse2015}.

This conceptual framework has been in place since the foundational
work of \citet{begelman1980}.  Yet debates continue on several fronts,
notably the physical processes responsible for shrinking the BH
separations below parsec-scales \citep[reviewed by][]{komossa2016}.
Finding candidate sub-parsec BBHs should help inform that debate, as
well as offer some assurance that these gravitational-wave progenitors
actually exist in Nature.  Direct imaging of BBHs on sub-parsec scales
is not yet possible, but such systems can be inferred from indirect
evidence like periodic modulation in the optical light curves
\citep[reviewed by][]{graham2016} or emission lines \citep[reviewed
  by][]{popovic2012} (but see \citet{liuj2016}), or oscillatory
structure in parsec-scale radio jets \citep[][and references therein]{mohan2016}.

From a systematic search for periodically-varying sources,
\citet{liu2015} reported a 1.5-yr modulation of the optical light
curves of PSO\,J334.2028+01.4075 (hereafter, PSO\,J334+01), a quasar
at $z = 2.060$ \citep{becker2001}.  \citet{liu2015} used the C\,IV
line from the spectrum \citep{becker2001} to estimate a quasar mass of
about 10 billion solar masses for $H_0 = 70$ km s$^{-1}$ Mpc$^{-1}$,
$\Omega_M$ = 0.3, and $\Omega_\Lambda$ = 0.7.  They assumed a scenario
wherein the mass accretion rate - thus the quasar luminosity - is
modulated by the binary's orbital motion.  Then the quasar mass plus
the rest-frame period of 0.5 yr suggested a BBH with a separation of
0.006 pc, corresponding to 0.7 $\mu$arcsec \citep{wright2006}.
However, analyses of light curves spanning longer times now disfavor
the persistence of the reported optical periodicity \citep{graham2015b,liu2016}.

Further analyses of optical light curves will no doubt ensue and
perhaps also explore other scenarios that could cause periodicity,
such as a precessing jet, a warped accretion disk or relativistic
Doppler boosting \citep{graham2016}.  In the interim, here we
investigate if PSO\,J334+01's radio continuum traits could provide
independent clues about a putative BBH.  We also set out to improve
the quasar's mass estimate.  We describe the new optical and radio
data in \S\ref{sec:obs_data}, discuss their implications in
\S\ref{sec:implications}, and close with a summary and conclusions in
\S\ref{sec:summary}.

\section{Observations, Data Processing, and Imaging}\label{sec:obs_data}

\subsection{Keck}\label{sec:keck}

\begin{figure*}
\centering
   \begin{overpic}[scale=0.42]{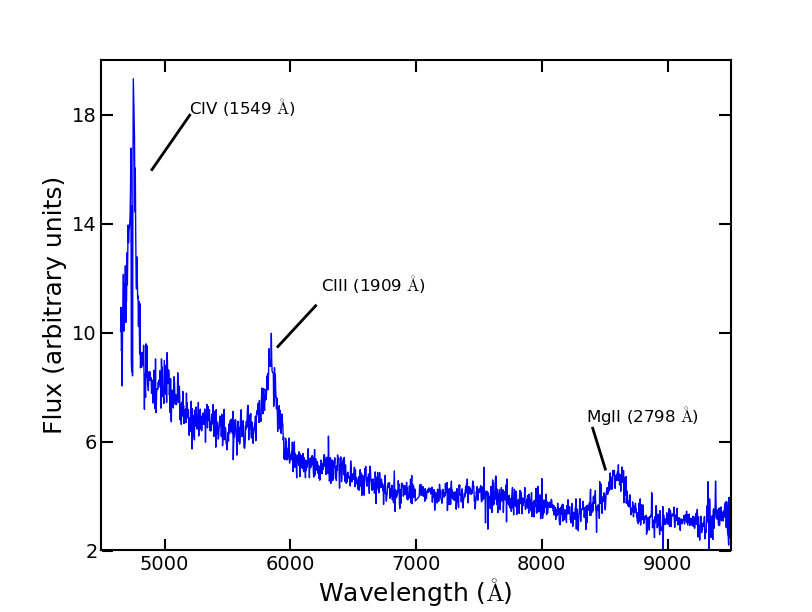}
     \put(58,37){\includegraphics[scale=0.6]{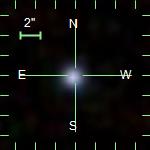}}
   \end{overpic}
\includegraphics[width=3.5in]{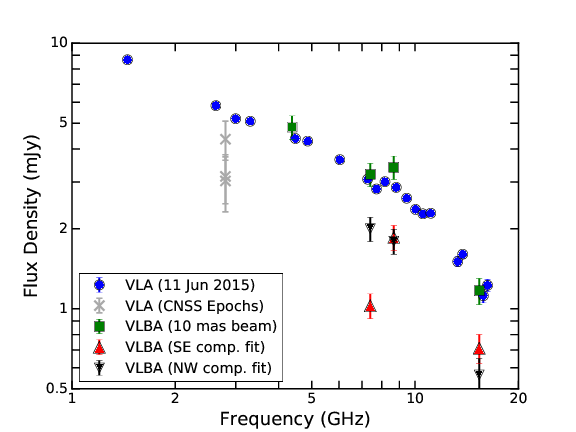}
\caption{\it Optical (left) and radio continuum (right) spectra of
  PSO\,J334+01 in the observed frame.  An optical image cutout of the
  quasar from the SDSS is shown as inset in the left panel.}
\label{fig:spec}
\end{figure*}

Since \cite{liu2015} used only a figure of an optical spectrum to
estimate the quasar's mass from an emission-line width, we obtained a
spectrum of PSO\,J334+01 with the DEIMOS instrument \citep{faber2003}
on the Keck II telescope on 2015 June 19.  The Keck spectrum, shown in
Figure~\ref{fig:spec}, reveals broad emission lines and no obvious
absorption features.  We fit the C\,III] (1909 $\mbox{\AA}$) and Mg\,II (2798
  $\mbox{\AA}$) emission lines with Gaussian profiles, and obtained
  observer-frame center wavelegths of 8592$\pm$8 $\mbox{\AA}$ and 5843$\pm$3
  $\mbox{\AA}$ respectively.  We therefore infer a redshift of
  2.063$\pm$0.001, and subsequent estimates for the rest-frame
  properties of PSO\,J334+01 will adopt that redshift.  We find that
  the Mg\,II line has a rest-frame FWHM of 5750$\pm$900 km s$^{-1}$.
  Our redshift and line-width from Keck data agree with those recently
  reported from lower-resolution Gemini data \citep{liu2016}.

\subsection{VLA}\label{sec:vla}

\begin{figure*}
\centering
\includegraphics[width=7.2in]{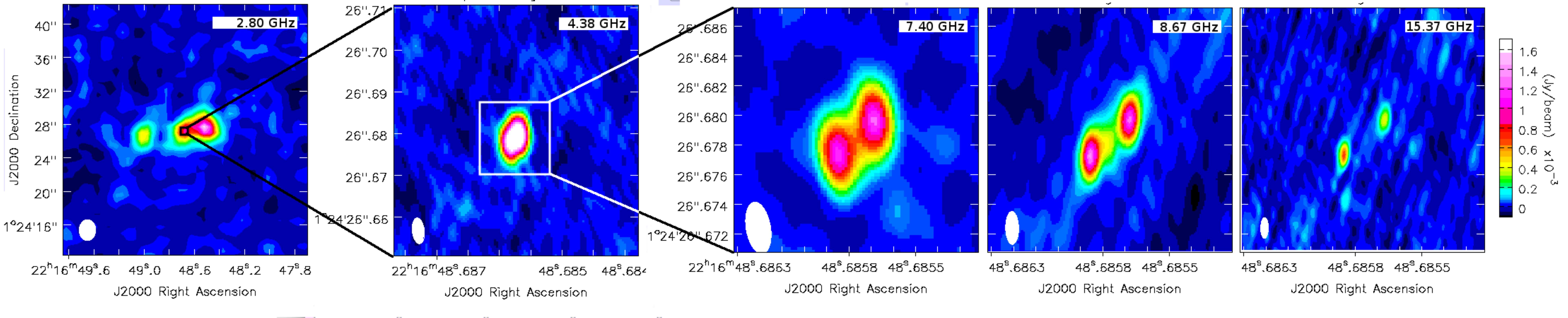}
\caption{\it VLA image cutout from the CNSS (far left) and VLBA image
  cutouts.  The three cutouts on the right are 16 mas $\times$ 16 mas
  in size.  The white ellipses show the angular resolutions at FWHM.
  The labels give the observed frequencies.  The colorbar on the right
  runs from 0.0 mJy beam$^{-1}$ (black) to 1.6 mJy beam$^{-1}$
  (white), with red representing $\sim$0.8 mJy beam$^{-1}$.  At the
  Keck redshift of PSO\,J334+01 the scale is 1 mas $=$ 8.3 pc.}
\label{fig:cutouts}
\end{figure*}

Our VLA observations of PSO\,J334+01 were made as part of the
Caltech-NRAO Stripe 82 Survey \citep[CNSS;][]{mooley2016} in the S
band centered at an observing frequency $\nu_o =$ 2.80 GHz, between
2013 December 21 and 2014 March 4.  The B configuration was used and
provided a FWHM angular resolution $\theta =$ 2.5 arcsec.  The CNSS
data were processed using a custom-developed AIPSLite pipeline,
followed by imaging in CASA \citep{mcmullin2007}.
Figure~\ref{fig:cutouts} shows the CNSS image cutout for which the
integrated and peak flux densities are $S_{\rm i} = 14.5\pm0.3$ mJy
and $S_{\rm p} = 3.6\pm0.3$ mJy, respectively.  Because the VLA peak
is point-like, the latter quantity is expressed in units of mJy.

We also obtained a VLA spectrum spanning $\nu_o =$ 1.45--16.26 GHz on
2015 June 11 during the move between the BnA and A configurations,
providing FWHM angular resolutions $\theta =$ 0.3--2.1 arcsec.  Those
data were processed using the NRAO CASA Pipeline and imaged manually
in CASA.  The point-like flux densities $S_{\rm p}$, with their
fitting uncertainties and 10\% amplitude calibration uncertainties
added in quadrature, appear in Table~\ref{tab:radio_measurements} and
Figure~\ref{fig:spec}.

\begin{table}
\centering
\caption{VLA and VLBA Flux Densities}
\label{tab:radio_measurements}
\begin{tabular}{rll} 
\hline\hline
$\nu_o$ & $S$   & $\sigma_S$\\   
(GHz)     & (mJy) & (mJy)\\
\hline
\multicolumn{3}{c}{VLA ($S_{\rm p}$}\\
\multicolumn{3}{c}{on 2015 Jun 11)}\\
\hline
 1.45 & 8.66 & 0.14\\
 2.63 & 5.82 & 0.13\\
 3.00 & 5.20 & 0.11\\
 3.31 & 5.08 & 0.07\\
 4.48 & 4.38 & 0.08\\
 4.86 & 4.28 & 0.08\\
 6.04 & 3.64 & 0.07\\
 7.27 & 3.08 & 0.06\\
 7.72 & 2.83 & 0.07\\
 8.17 & 3.01 & 0.06\\
 8.81 & 2.87 & 0.06\\
 9.45 & 2.61 & 0.06\\
10.03 & 2.37 & 0.07\\
10.54 & 2.28 & 0.07\\
11.11 & 2.29 & 0.06\\
13.31 & 1.51 & 0.07\\
13.76 & 1.61 & 0.06\\
15.81 & 1.12 & 0.07\\
16.26 & 1.23 & 0.06\\
\hline
\multicolumn{3}{c}{VLBA ($S_{\rm i}$ after}\\
\multicolumn{3}{c}{convolving with}\\
\multicolumn{3}{c}{a 10 mas beam)}\\
\hline
4.38  & 4.83 & 0.49\\
7.40  & 3.21 & 0.33\\
8.67  & 3.41 & 0.34\\
15.37 & 1.17 & 0.13\\
\hline
\end{tabular}
\end{table}

At $\nu_o =$ 2.80 GHz the point-like VLA source at matched resolutions (2'')
has $S_{\rm p} =$ 3.15$\pm$0.66 mJy, 4.35$\pm$0.71 mJy, 3.03$\pm$0.73
mJy, and 1.70$\pm$0.76 mJy following 0d, 23d, 49d, and 73d after 2013
December 21, respectively \citep{mooley2016}.  (The CNSS image in
Figure~\ref{fig:cutouts} is a stack of the first three epochs.)  At
the same frequency the single-epoch VLA spectrum at 567d shows $S_{\rm
  p} =$ 5.51$\pm$0.17 mJy.  This last datum was acquired at a higher
angular resolution, so its record-high value cannot be attributed to
contamination from lobe emission and must represent true variability.

\subsection{VLBA}\label{sec:vlba}

The VLBA observations of PSO\,J334+01 were obtained under the proposal
code BM438.  In order to maximize the bandwidth and sensitivity, we
observed in the C, X, and U bands using the DDC personality of the
Roach Digital Backend.  Our setup included four intermediate
frequencies (IFs) having 256 channels each to achieve a 128 MHz
bandwidth per IF, dual polarizations, and 2 bits per sample.  The
center frequencies of the IFs were chosen to avoid regions of known
radio-frequency interference.  Accordingly, we used lower sidebands
for the setup at C band (centered on $\nu_o =$ 4.38 GHz and 7.40 GHz),
and upper sidebands for the setups at X band (centered on $\nu_o =$
8.60 GHz and 8.73 GHz) and U band (centered on $\nu_o =$ 15.30 GHz and
15.43 GHz).  Due to the limited LST range accessible for observing
equatorial sources with the VLBA, we split the schedules for each band
into two blocks of 2.5 hours each.  To mitigate the effects of time
variability, each band's scheduling blocks were observed within about
two days.  We used J2253+1608 for fringe finding and J2217+0220 for
phase calibration.  The log of observations is given in
Table~\ref{tab:vlba_log}.

%
%

\begin{table}
\centering
\caption{VLBA Observing Log}
\label{tab:vlba_log}
\begin{tabular}{lp{1.47cm}ll}
\hline\hline 
Sched. & UT Start & UT Start & Band \\
Block  & Date     & Time     & \\
\hline
1     & 2016 Mar 24 & 15:32 & X\\
2     & 2016 Mar 26 & 16:24 & X\\
3     & 2016 Mar 30 & 15:08 & C\\
4     & 2016 Apr 01 & 15:30 & C\\
5     & 2016 May 12 & 12:19 & U\\
6     & 2016 May 14 & 12:41 & U\\
\hline
\end{tabular}
\end{table}

For calibrating the VLBA data, we implemented the recipe from Chapter
9 of the AIPS cookbook in AIPS \citep{griesen2003} accessed via
ParselTongue.  In brief, we imported the FITS IDI data using the FITLD
task with DIGICOR and WTTHRESH parameters set to 1 and 0.7
respectively.  We then flagged (i) the data having antenna elevation
less than 15$\deg$, (ii) first 4 seconds of each scan, and (iii) 10
channels towards the beginnings and ends of the IFs, where the
bandpass is usually poorly characterized.  We then derived and applied
the parallactic angle correction using CLCOR, followed by the digital
sampler bias corrections using ACCOR and CLCAL.  For instrumental
phase corrections, we used PCOR and CLCAL.  Thereafter, we derived
continuum amplitude calibration using ACSCL and APCAL tasks, derived
the bandpass using BPASS with the fringe finder as the calibrator, and
after each task applied the calibration using CLCAL.  To remove
residual phase errors in the data, we used FRING.  Finally, we applied
the calibration tables to the target data with SPLIT and wrote the
calibrated data to disk using FITTP.

We imported the calibrated UVFITS data for PSO\,J334+01 into CASA with
{\tt importuvfits} and performed interactive imaging using {\tt
  clean}.  We imaged each IF for the C band observation separately,
and combined both IFs for X band and U band while imaging.  The pixel
size was chosen appropriately to sample the PSF with five or more
pixels, and the image size was set to 1024$\times$1024 pixels.
Figure~\ref{fig:cutouts} shows the VLBA image cutouts at $\nu_o =$
4.38 GHz (C band IF 1), $\nu_o =$ 7.40 GHz (C band IF2), $\nu_o =$
8.67 GHz (X band), and $\nu_o =$ 15.37 GHz (U band).
Table~\ref{tab:img_parms} gives the native resolution and RMS noise
achieved in each image.  Additionally, we fit the components in each
image with Gaussian components using SAD in AIPS, and the resultant
parameters appear in Table~\ref{tab:img_parms} and
Figure~\ref{fig:spec}.  At the frequencies where two VLBA components,
SE to the southeast and NW to the northwest, can be discerned, they
are separated by 3.6 mas along an elongation position angle (PA),
measured north through east, of 139$\deg$.  The SE component appears
to be somewhat more compact than the NW one.

\begin{table*}
\centering
\tiny
\caption{VLBA Image Properties and Source Fits}
\label{tab:img_parms}
\begin{tabular}{rlllllllllllr} 
\hline\hline $\nu_o =$ & BMAJ& BMIN& BPA& RMS& Component& $S_{\rm p}$ & $S_{\rm i}$&
RA Offset & Dec Offset & Maj & Min & PA\\ 
(GHz)& (mas)& (mas)& (deg)& ($\mu$Jy beam$^{-1}$)& & (mJy beam$^{-1}$)& (mJy)&
(mas)& (mas)& (mas)& (mas)& (deg)\\ 
\hline 
4.38 & 6.2 & 2.9 & 6.7 & 25 & & $2.730\pm0.025$ & $4.687\pm0.063$ &
$11.17\pm0.02$ & $-0.94\pm0.03$ & $7.08\pm0.06$ & $4.38\pm0.04$ &
$164\pm1$\\
\hline 
7.40 & 3.5 & 1.6 & 9.5 & 23 & SE & $1.075\pm0.022$ & $1.028\pm0.037$ &
$12.88\pm0.01$ & $-2.63\pm0.03$ & $3.64\pm0.08$ & $1.53\pm0.03$ &
$10\pm1$\\
& & & & & NW & $1.114\pm0.021$ & $2.001\pm0.055$ & $10.55\pm0.02$ &
$-0.73\pm0.03$ & $4.37\pm0.08$ & $2.40\pm0.05$ & $167\pm1$\\
\hline
8.67 & 2.2 & 0.9 & 1.0 & 29 & SE & $1.296\pm0.031$ & $1.858\pm0.068$ &
$12.75\pm0.01$ & $-2.69\pm0.03$ & $2.54\pm0.06$ & $1.11\pm0.03$ &
$173\pm1$\\
& & & & & NW & $1.213\pm0.031$ & $1.798\pm0.070$ & $10.21\pm0.01$ &
$-0.29\pm0.03$ & $2.50\pm0.07$ & $1.14\pm0.03$ & $173\pm1$\\
\hline 
15.37 & 1.4 & 0.5 & -0.8 & 30 & SE & $0.692\pm0.031$ & $0.711\pm0.054$
& $12.76\pm0.01$ & $-2.65\pm0.03$ & $1.42\pm0.06$ & $0.54\pm0.02$ &
$173\pm1$\\
& & & & & NW & $0.395\pm0.029$ & $0.564\pm0.065$ & $10.03\pm0.02$ &
$-0.29\pm0.05$ & $1.61\pm0.12$ & $0.66\pm0.05$ & $172\pm3$\\
\hline
\multicolumn{13}{p{6.5in}}{Notes: (1) BMAJ, BMIN and BPA are the
  parameters of the synthesized beam at the image's native FWHM
  resolution.  (2) RMS is the image's root-mean-square noise.  (3)
  Identifies the component to the southeast (SE) or the northwest (NW)
  in Figure~2.  (4) $S_{\rm p}$ and $S_{\rm i}$ are the peak and
  integrated flux densities, respectively.  The quoted uncertainties
  are statistical-only.  There is an additional standard amplitude
  calibration uncertainty of $\sim$10\%.  (5) RA Offset and Dec Offset
  are relative to $\alpha(J2000) =$ 22h16m48.6851s and
  $\delta(J2000)=$ +01$^o$24'26.681''.  (6) Maj, Min and PA are the
  deconvolved major axis, minor axis, and PA respectively of the
  Gaussian fitted to the component.}
\end{tabular}
\end{table*}

In order to better compare the VLA flux densities with the VLBA ones,
we convolved each VLBA image with a circular 10 mas beam, and used SAD
to obtain integrated flux densities $S_{\rm i}$.  Those flux
densities, with their fitting uncertainties and 10\% amplitude
calibration uncertainties added in quadrature, appear in
Table~\ref{tab:radio_measurements} and Figure~\ref{fig:spec}.

\begin{figure}
\centering
\includegraphics[width=3.4in]{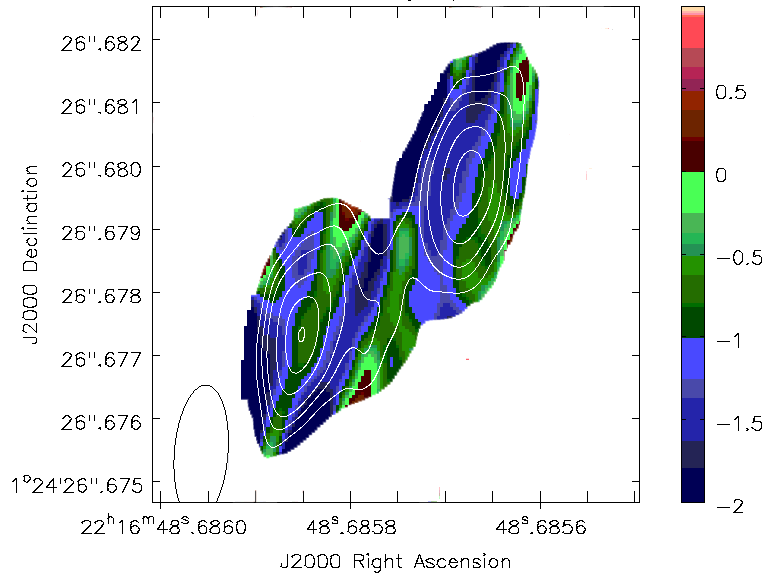}
\caption{\it VLBA spectral index image of PSO\,J334+01 between $\nu_o
  =$ 8.67 GHz and 15.37 GHz.  The contours are 10, 15, 20, 30, 40,
  50~$\times$20 $\mu$Jy at 12.02 GHz.  Regions fainter than
  5$\times$20 $\mu$Jy beam$^{-1}$ are blanked.  The ellipse shows the angular
  resolution at FWHM.  At the Keck redshift of PSO\,J334+01 the scale
  is 1 mas $=$ 8.3 pc.}
\label{fig:alpha}
\end{figure}

Since there is significant extended structure on VLBA scales, we
generated a spectral index image between $\nu_o =$ 8.67 GHz and 15.37
GHz.  To create this image, we combined the VLBA X-band and U-band
data using {\tt concat} in CASA, and then imaged using the {\tt clean}
task with two Taylor terms, 512$\times$512 pixels and a pixel size of
0.05 mas.  The emission is close to the VLBA pointing center so no
wide-band primary beam correction was necessary.  We define the
spectral index $\alpha$ according to $S\propto\nu^\alpha$.  The VLBA
spectral index image appears in Figure~\ref{fig:alpha}, overlayed with
contours from the associated image at the center frequency of $\nu_o =$
12.02 GHz.  The uncertainties in the spectral indices near the higher
contours are $\leqslant$0.25, while those near the lower contours are
$\leqslant$0.55.  The SE component appears to have a somewhat flatter
spectrum ($\alpha_{\rm SE} \sim -0.6\pm0.2$) than that of the NW
component ($\alpha_{\rm NW} \sim -1.1\pm0.1$).

\section{Implications}\label{sec:implications}

A VLA image of PSO\,J334+01 at $\nu_o =$ 1.4 GHz with a FWHM
resolution of 5 arcsec (42 kpc) showed it was partially resolved
\citep{becker1995}.  The kiloparsec-scale structure is better revealed
in our CNSS image (Figure~\ref{fig:cutouts}) at $\nu_o =$ 2.80 GHz
with a FWHM resolution of 2.5 arcsec (21 kpc).  Our image reveals a
point-like VLA source that is straddled by asymmetric lobes that span
8 arcsec (66 kpc) and define an elongation PA of 100$\deg$.  At this
rest-frame frequency of $\nu_r =$ 8.58 GHz, the ratio of point-like to
extended flux densities is $R = 0.33\pm0.03$.  This leads us to follow
\citet{hough2002} and \citet{hough2008}, and classify PSO\,J334+01 as
a lobe-dominated quasar, although we recognize that those authors
adopted a lower $\nu_r =$ 5 GHz for their 25-quasar sample.  $R$ may
be thought of as quantifying the relative strength of
relativistically-beamed features, if any, normalized to the unbeamed
extended emission.  Those authors favor $R$ as an indicator of the
hypothetical orientation angle of a lobe-dominated quasar relative to
our line-of-sight, $\theta_{\rm los}$, and suggest $\theta_{\rm los}
\sim 10\deg - 40\deg$ for their sample.
The CNSS image also reveals that the western lobe dominates over the eastern lobe.
The integrated flux densities of these two lobes at $\nu_o =$ 3.00 GHz (on 2015 June 11) 
are $7.43\pm0.26$ mJy and $3.03\pm0.23$ mJy respectively.

From our VLA spectrum of PSO\,J334+01 on 2015 June 11
(Figure~\ref{fig:spec}), the spectral index between $\nu_o =$ 1.45 GHz
and 8.17 GHz ($\nu_r =$ 4.44 GHz and 25.02 GHz) is $\alpha = -0.61$,
while that between $\nu_o =$ 8.17 GHz and 16.26 GHz ($\nu_r =$ 25.02
GHz and 49.80 GHz) is $\alpha = -1.30$.  The uncertainty in each
spectral index is $<$0.1.  For synchrotron emitting sources, a
spectrum curving downward at higher frequencies is expected because
the highest-energy electrons radiate away their energy the most
rapidly (the emission frequency and rate of energy loss are both proportional to the square of the energy).
This results in a paucity of electrons emitting at the high-frequency
end if there is no replenishment of the high-energy electrons.  Also,
Figure~\ref{fig:spec} shows that the VLA spectrum matches that from
the convolved VLBA images, suggesting that the majority of the VLA
point-like emission originates on the parsec scales probed by the VLBA.

Table~\ref{tab:img_parms} and Figure~\ref{fig:spec} imply that the
VLBA component to the SE has an inverted spectrum between $\nu_o =$
7.40 GHz and 8.67 GHz ($\nu_r =$ 22.67 GHz and 26.56 GHz), with an
apparent $\alpha = 3.7\pm1.0$ in the rest frame.  While it is possible
that some of the emission of the SE component at $\nu_o =$ 7.40 GHz
has not been accounted for in the Gaussian fit, the inverted spectrum
is likely real.  This strongly suggests that the SE component is
synchrotron self-absorbed and represents the VLBA ``core'' that has
ejected the NW component.  The SE component appears to be more compact
than the NW one (Table~\ref{tab:img_parms}) and has a flatter spectrum
than the NW one (Figure~\ref{fig:alpha}), lending additional support
to this hypothesis. It is worth noting that the one-sided parsec-scale jet (the NW 
component) points toward the brighter (western) lobe seen on kiloparsec scales in the VLA images.
The attributes of components SE and NW at $\nu_o
=$ 15.37 GHz (Table~\ref{tab:img_parms}) imply rest-frame brightness
temperatures of 15 Million K and 9 Million K, respectively.  Such
values are high enough to imply nonthermal emission, but not so high
as to exceed the equipartition brightness temperature or the inverse
Compton limit for a non-moving component.


Like \cite{hough2002}, we quantify a fractional variability amplitude
as $f_{\rm v} = (S_{\rm max} - S_{\rm min})/S_{\rm min}$.  From the
VLA monitoring of PSO\,J334+01 at $\nu_r =$ 8.58 GHz on timescales of
weeks to months to 1.6 yr (\S\ref{sec:vla}), we estimate $f_{\rm v} =
2.2\pm1.0$.  This value is poorly constrained but resembles those for
the more variable VLBI ``cores'' at $\nu_o =$ 8.4 or 10.7 GHz in the
\cite{hough2002} sample.  Of the 14 sources studied, 5 show no
evidence for fractional variability amplitudes above 0.1, while the
remaining 9 sources show $f_{\rm v} =$ 0.12--2.6.  A caveat is that we
do not have $f_{\rm v}$ information for the SE component in
PSO\,J334+01, its supposed VLBA ``core''.  \citet{hough2002} also
reported a strong correlation between $f_{\rm v}$ and the orientation
indicator $R$.  For PSO\,J334+01, $R = 0.33\pm0.03$ (or log$R$ =
-0.48$\pm$0.04) and it would adhere to the correlation only if $f_{\rm
  v} \lesssim 1.5$.  Monitoring PSO\,J334+01 with improved accuracy
could test this expectation of modest amplitude variability.

The Mg\,II line width (\S\ref{sec:keck}) can be used to estimate the
black hole mass of PSO\,J334+01.  We follow the procedure in
\citet{liu2016} and employ equation A6 of \citet{mclure2004}.  Since we
do not have a flux-calibrated optical spectrum, we take the 3000 $\mbox{\AA}$
continuum flux from \citet{liu2016}.  We thus obtain a black hole mass
of log(M$_{\rm BH}$/M$_\odot$) $= 9.1\pm0.3$ (error statistical-only),
in agreement with \citet{liu2016}. We also find that the Mg\,II (2798$\mbox{\AA}$) 
emitted line FWHM ($23\pm4$ $\mbox{\AA}$) is in agreement with the anticorrelation 
between log$R$ and Mg\,II emitted line width reported by \cite{hough2002} and 
\cite{aars2005}.

Amongst candidate sub-parsec BBHs discovered by means of periodic
optical signals, only OJ\,287 \citep[][and references
  therein]{valtonen2012}, and PG\,1302-102
\citep{graham2015,kun2015,mohan2016} have been examined with the VLBA.
Systems that can be studied via these two independent means have the
potential to be especially strong BBH contenders.  If a radio jet is
launched from one BH member of a binary, then its structure could
exhibit wiggles caused by the orbital motion of the BH and/or a
helical pattern with a conical geometry caused by precession of the
spin axis of the BH \citep{gower1982,kaastra1992,kun2014}.

The persistence of the optical periodicity of PSO\,J334+01 is now disfavored, but the
quasar's radio continuum traits can still offer clues about a putative
BBH.  For PSO\,J334+01, its VLBA components SE and NW are separated by
3.6 mas (30 pc) and define an elongation PA\footnote{The PA values may 
reflect amplification by projection of smaller intrinsic bends.} of 139$\deg$
(\S\ref{sec:vlba}), markedly different from the elongation PA of
100$\deg$ measured on VLA or kiloparsec scales.  A twist of $39\deg$
between elongation PAs on parsec- and kiloparsec-scales is atypically
large for a lobe-dominated quasar, being achieved by only one of the
16 quasars with such measurements available in the \citet{hough2002} sample: 3C\,207 which twists by
$40\deg$.  Both 3C\,207 and PSO\,J334+01 have similar ratios of
point-like to extended flux densities, $R \sim 0.49$ \citep{hough2002}
and $R = 0.33$, respectively, so both could be near the extreme of
$\theta_{\rm los} \sim 10\deg$ postulated for this quasar class.

Extensive VLBI monitoring of 3C\,207 underpins a model wherein jet
components are ejected at different angles around a cone, with the
flow then recollimating downstream to align with the kiloparsec-scale
emission \citep{hough2013}.  As noted above, a natural way to develop
an inner cone is to have the jet-producing BH precess due to the
binary's orbital motion \citep{gower1982,kaastra1992}.  This suggests
that VLBA monitoring of PSO\,J334+01 could be productive and, perhaps,
show evidence for an inner-cone configuration.  Intriguingly,
\cite{kulkarni2016} describe how particularly-wide cone angles could
develop as energy loss via gravitational waves causes a BBH orbit to
shrink in size, shorten in period, and widen in (intrinsic) cone
angle.

The structure of the radio lobes seen in PSO\,J334+01 can also be discussed in the context of X-shaped radio galaxies \citep[XRGs; e.g.][]{cheung2007}, a subset which may be 
signposts of SMBH coalescence \citep[e.g.][]{roberts2016}.
\cite{roberts2015} and \cite{roberts2016} suggest that XRGs with distortions near the inner ends of the on-axis lobes and those having inversion-symmetric structures 
may likely be formed due to back-flow deflections by the thermal halos of the host ellipticals or due to swings/flips of the jet axis during a galaxy merger event.
They also suggest that the XRGs having distortions near the outer ends of the on-axis lobes are indicative of axis precession, which may be caused by BBHs.
PSO\,J334+01 represents this latter (BBH) case, albeit the putative structure connecting the inner (parsec-scale) lobes with the outer (kiloparsec-scale) lobes remains undetected in our observations.

\section{Summary and Conclusion}\label{sec:summary}

The optical-continuum periodicity in PSO\,J334.2028+01.4075\footnote{The optical periodicity is now disfavored by \cite{liu2016}, but it may still be consistent with a BBH, 
if the light curve is more complicated than a simple sinusoid, or if there is a change in period due to orbital decay.}
led \cite{liu2015} to propose the presence of a binary black hole (BBH) with an orbital separation of about 0.006 pc (0.7 $\mu$arcsec) in the quasar.
Motivated by the need for independent clues about a putative BBH, we used the VLBA and VLA to study the parsec- and kiloparsec-scale radio emission of the quasar.
We additionally obtained an optical spectrum (Figure~\ref{fig:spec}), which reveals broad emission lines and no obvious absorption features, of the quasar and inferred a redshift of 2.063$\pm$0.001.
From the Mg\,II (2798$\mbox{\AA}$) line width, we estimated the black hole mass of PSO\,J334+01 as log(M$_{\rm BH}$/M$_\odot$) $= 9.1\pm0.3$.
Both, the redshift and the black hole mass, are in agreement with \citet{liu2016}.

The VLA image (Figure~\ref{fig:cutouts}) at an observed frequency of 2.80 GHz reveals a point-like source straddled by asymmetric lobes that span 8 arcsec (66 kpc).
The ratio of point-like to extended flux densities of $0.33\pm0.03$ (at a rest-frame frequency of 8.58 GHz) led us to classify PSO\,J334+01 as a lobe-dominated quasar.
Flux density measurements between six epochs observed over 1.5 years reveals modest amplitude variability in PSO\,J334+01.

Figure~\ref{fig:cutouts} shows the VLBA image cutouts between observed frequencies of 4.38 GHz and 15.37 GHz.
These high-resolution radio images reveal two components, one to the southeast or SE and another to the northwest or NW, separated by 3.6 mas (30 pc).
The VLA and VLBA images of PSO\,J334+01 together reveal an atypically large twist of $39\deg$ between the elongation position angles of the lobes on parsec and kiloparsec scales.
The SE component in the VLBA images appears to be somewhat more compact and having a flatter spectrum than the NW component.
We tentatively identify the former component as the VLBA ``core'' from which the latter was ejected.
The VLA follow-up spectrum between 4--16 GHz (in the observer frame; Figure~\ref{fig:spec}) matches that from the convolved VLBA images, suggesting that the majority of the VLA point-like emission 
originates on the parsec scales probed by the VLBA.

The rare twist between the lobes of PSO\,J334+01 on parsec- and kiloparsec-scales may indicate that the quasar is ejecting jet components over an inner cone that traces 
a precessing jet in a BBH system.
We advocate further VLBA monitoring of the quasar, which could perhaps show evidence for an inner-cone configuration.
All in all, the twisted radio structure implies that PSO\,J334+01 remains a supermassive binary black hole candidate, despite the optical-continuum periodicity being disfavored by \cite{liu2016}.

{\it Acknowledgements: We are grateful to David Hough for his insightful comments on the manuscript.
We thank the anonymous referee for the helpful suggestions.
KPM's research is supported by the Oxford
  Centre for Astrophysical Surveys which is funded through generous
  support from the Hintze Family Charitable Foundation.  The National
  Radio Astronomy Observatory is a facility of the National Science
  Foundation operated under cooperative agreement by Associated
  Universities, Inc.  Some of the data presented herein were obtained
  at the W.M. Keck Observatory, which is operated as a scientific
  partnership among the California Institute of Technology, the
  University of California and the National Aeronautics and Space
  Administration. The Observatory was made possible by the generous
  financial support of the W.M. Keck Foundation.  The authors wish to
  recognize and acknowledge the very significant cultural role and
  reverence that the summit of Mauna Kea has always had within the
  indigenous Hawaiian community.  We are most fortunate to have the
  opportunity to conduct observations from this mountain.  We thank
  the NRAO staff for scheduling the VLBA and VLA observations, and the
  Keck staff for executing the optical observations.  Any opinions,
  findings, and conclusions or recommendations expressed in this
  material are those of the authors and do not necessarily reflect the
  views of the National Science Foundation.  }



\begin{thebibliography}{99}

\bibitem[Aars et al. (2005)]{aars2005} Aars, C. E., Hough, D. H., Yu,
L. H., et al. 2005, AJ, 130, 23

\bibitem[Barausse et al. (2015)]{barausse2015} Barausse, E.,
Bellovary, J., Berit, E., et al. 2015, Journal of Physics:
Conference Series, 610, 012001

\bibitem[Becker, White \& Helfand (1995)]{becker1995} Becker, R. H.,
White, R. L., \& Helfand, D. 1995, ApJ, 450, 559

\bibitem[Becker et al. (2001)]{becker2001} Becker, R. H., White,
R. L., \& Gregg, M. D., et al. 2001, ApJS, 135, 227

\bibitem[Begelman, Blandford \& Rees (1980)]{begelman1980} Begelman,
M. C., Blandford, R. D., \& Rees, M. J. 1980, Nature, 287, 307

\bibitem[Burke-Spolaor (2015)]{burke-spolaor2015} Burke-Spolaor,
S. 2015, arXiv:1511.07869

\bibitem[Cheung(2007)]{cheung2007} 
Cheung C. 2007, ApJ, 33, 2097

\bibitem[Colpi (2014)]{colpi2014} Colpi, M. 2014, Space Science
Reviews, 183, 189

\bibitem[Comerford et al. (2015)]{comerford2015} Comerford, J. M.,
Pooley, D., Barrows, R. S., et al. 2015, ApJ, 806, 219

\bibitem[Deane et al. (2014)]{deane2014} Deane, R. P., Paragi, Z.,
Jarvis, M. J., et al. 2014, Nature, 511, 57

\bibitem[D'Orazio, Haiman \& Schiminovich (2015)]{dorazio2015}
  D'Orazio, D. J., Haiman, Z., \& Schiminovich, D. 2015, Nature, 525,
  351

\bibitem[Faber et al. (2003)]{faber2003} Faber, S. M., Phillips,
A. C., Kibrick, R. I., et al. 2003, SPIE, 4841, 1657.

\bibitem[Fu et al. (2015a)]{fu2015a} Fu, H., Myers, A. D., Djorgovski,
S. G., et al. 2015a, ApJ, 799, 72

\bibitem[Fu et al. (2015b)]{fu2015b} Fu, H., Wrobel, J. M., Myers,
A. D., et al. 2015b, ApJL, 815, L6

\bibitem[Gower et al. (1982)]{gower1982} Gower, A. C., Gregory, P. C.,
Unruh, W. G., \& Hutchings, J. B. 1982, ApJ, 262, 478

\bibitem[Graham et al. (2015a)]{graham2015} Graham, M. J.,
Djorgovski, S. G., Stern, D., et al. 2015a, Nature, 518, 74

\bibitem[Graham et al. (2015b)]{graham2015b} Graham, M. J.,
Djorgovski, S. G., Stern, D., et al. 2015b, MNRAS, 453, 1562

\bibitem[Graham et al. (2016)]{graham2016} Graham, M. J., Djorgovski,
S. G., Stern, D., et al. 2016, arXiv:1612.07271
  
\bibitem[Griesen (2003)]{griesen2003} Griesen, E. W. 2003, in
Information Handling in Astronomy, ed. A. Heck (Dordrecht: Kluwer),
109

\bibitem[Hough et al. (2002)]{hough2002} Hough, D. H., Vermeulen,
R. C., Readhead, A. C. S., Cross, L. L., Barth, E. L. et al. 2002,
AJ, 123, 1258

\bibitem[Hough (2008)]{hough2008} Hough, D. H. 2008, in ASP
Conf. Ser. 386, Extragalactic Jets: Theory and Observation from
Radio to Gamma Ray, ed. T. A. Rector \& D. S. De Young, (San
Francisco, CA: ASP), 274

\bibitem[Hough (2013)]{hough2013} Hough, D. H. 2013, in EPJ Wb of
Conferences, 61, 08009

\bibitem[Kaastra \& Roos(1992)]{kaastra1992} Kaastra, J. S., \& Roos,
N. 1992, A\&A, 254, 96

\bibitem[Komossa, Baker \& Liu (2015)]{komossa2016} Komossa, S.,
Baker, J. G., \& Liu, F. K. 2016, Astronomy in Focus, Proceedings of the IAU, Volume 29B, 292

\bibitem[Kulkarni \& Loeb (2016)]{kulkarni2016} 
Kulkarni, G. \& Loeb, A. 2016, MNRAS, 456, 3964
  
\bibitem[Kun et al. (2014)]{kun2014} Kun, E., Gabanyi, K. E.,
Karouzos, M., et al. 2014, MNRAS, 445, 1370

\bibitem[Kun et al. (2015)]{kun2015} Kun, E., Frey, S., Gabanyi,
K. E., et al. 2015, MNRAS, 454, 1290

\bibitem[Liu et al. (2015)]{liu2015} Liu, T., Gezari, S., Heinis, S.,
et al. 2015, ApJL, 803, L16

\bibitem[Liu, Eracleous \& Halpern (2016)]{liuj2016} Liu, J.,
Eracleous, M., \& Halpern, J. P. 2016, ApJ, 817, 42

\bibitem[Liu et al. (2016)]{liu2016} Liu, T., Gezari, S., Burgett, W.,
Chambers, K., Draper, P. et al. 2016, ApJ, 833, 6

\bibitem[McLure \& Dunlop (2004))]{mclure2004} McLure, R. J., \&
Dunlop, J. S. 2004, MNRAS, 352, 1390

\bibitem[McMullin et al. (2007)]{mcmullin2007} McMullin, J. P., Waters,
B., Schiebel, D., et al. 2007, in ASP Conf. Ser. 376, Astronomical
Data Analysis Software and Systems, IVI, ed. R. A. Shaw, F. Hill, \& 
D. J. Bell (San Francisco, CA: ASP), 127

\bibitem[Mohan et al. (2016)]{mohan2016} Mohan, P., An, T., Frey, S.,
et al. 2016, MNRAS, 463, 1812

\bibitem[Mooley et al. (2016)]{mooley2016} Mooley, K. P., Hallinan,
G., Bourke, S., et al. 2016, ApJ, 818, 105

\bibitem[Popovic(2012)]{popovic2012} Popovic, L. C. 2012, New Astron.\
Rev., 56, 74

\bibitem[Roberts et al.(2015)]{roberts2015} 
Roberts, D., Cohen, J., Lu, J., Saripalli, L. \& Subrahmanyan, R. 2016, ApJS, 220, 7

\bibitem[Roberts, Saripalli \& Subrahmanyan(2016)]{roberts2016} 
Roberts, D., Saripalli, L. \& Subrahmanyan, R. 2016, ApJL, 810, L6
%
\bibitem[Schnittman (2013)]{schnittman2013} Schnittman, J. D. 2013,
Class. Quantum Grav., 20, 244007
%
\bibitem[Valtonen \& Wiik (2012)]{valtonen2012} Valtonen, M. J. \&
Wiik, K. 2012, MNRAS, 421, 1861

\bibitem[Volonteri, Miller \& Dotti (2009)]{volonteri2009}
Volonteri, M., Miller, J. M., \& Dotti, M. 2009, ApJL, 703, L86
%
%
\bibitem[Wright (2006)]{wright2006} Wright, E. L. 2006, PASP, 118,
1711


\end{thebibliography}
\end{document}